\documentclass[a4paper]{article}

\usepackage{INTERSPEECH2020}

\usepackage{subcaption}
\usepackage{tikz}
\usetikzlibrary{bayesnet}
\usetikzlibrary{decorations.pathreplacing}
\usetikzlibrary{calc}
\usepackage{pgfplots}
\usepackage{fontawesome}
\usepgfplotslibrary{fillbetween}

\pgfplotsset{compat=1.14}

\usepackage{mystyle}

\title{Bayesian Subspace HMM for the Zerospeech 2020 Challenge}
\name{Bolaji Yusuf$^{1,2}$, Lucas Ondel$^2$}
\address{
  $^1$Bogazici University, Department of Electrical and Electronics Engineering, Turkey\\
  $^2$Brno University of Technology, Speech@FIT and IT4I Center of Excellence, Brno, Czech Republic
  \thanks{The work was supported by Czech Ministry of Interior project No. VI20152020025 "DRAPAK", Czech Ministry of Education, Youth and Sports project No. LTAIN19087 "Multi-linguality in speech technologies", and by European Union’s Horizon 2020 project No. 870930 - WELCOME.}}
\email{bolaji.yusuf@boun.edu.tr, iondel@fit.vutbr.cz}
\newcommand\Lucas[1]{#1}

\begin{document}
\maketitle
\begin{abstract}
  In this paper we describe our submission to the Zerospeech 2020 challenge, where the participants are required to discover latent representations from unannotated speech, and to use those representations to perform speech synthesis, with synthesis quality used as a proxy metric for the unit quality. In our system, we use the Bayesian Subspace Hidden Markov Model (SHMM) for unit discovery. The SHMM models each unit as an HMM whose parameters are constrained to lie in a low dimensional subspace of the total parameter space which is trained to model phonetic variability. Our system compares favorably with the baseline on the human-evaluated character error rate while maintaining significantly lower unit bitrate.


\end{abstract}


\section{Introduction}
Learning useful unsupervised representations of data is one of the most important research questions of modern machine learning. In Automatic Speech Recognition (ASR) particularly, there is a wide performance schism between systems in resource-rich languages and low resource languages due to the dependence of contemporary ASR technology on large annotated speech corpora. There has therefore been growing interest in using unsupervised machine learning approaches to speech processing to bridge this gap. One line of research is Acoustic Unit Discovery (AUD) where the goal is to discover a set of units, similar to phones, from unlabeled speech in a language.

The overarching theme of most AUD methods is the postulation of a latent space which represents the units and which is inferred from the data. Bayesian approaches model the problem as a generative process such as an HMM or GMM with a Dirichlet process prior so that both the number of units and the parameters of those units' models can be inferred~\cite{lee2012nonparametric,ondel2016variational,kamper2017segmental,heck2017feature,ondel2018bayesian, ondel2017bayesian, Ondel2019shmm}. Another approach is the use of neural networks which usually learn continuous latent spaces rather than discrete units. Autoencoders have been used to this end, with further constraints such as speaker invariance costs or correspondence training~\cite{kamper2015unsupervised,kamper2019truly,yusuf2019temporally}. Siamese neural networks have also been applied, where the labels for training are obtained from an unsupervised term detection system~\cite{thiolliere2015hybrid,zeghidour2016deep}.

Recently, Bayesian-neural-network hybrids have become popular because they combine the modeling power of neural networks with the structured self-regularizing properties of Bayesian models. Variants of the Variational AutoEncoder (VAE) have been especially successful~\cite{kingma2013auto}. Among these are the vector quantized VAE~\cite{chorowski2019unsupervised,Tjandra2019,Eloff2019}, HMM-VAE~\cite{ebbers2017hidden, glarner2018full}.

The AUD system we adopt in this work is one such Bayesian hybrid, the Subspace Hidden Markov Model (SHMM) system~\cite{Ondel2019shmm}. It models each acoustic unit as an HMM whose parameters are constrained to be in a low-dimensional \textit{phonetic} subspace of the parameter space.

The Zerospeech challenge~\cite{versteegh2015zero, dunbar2017zero, dunbar2019zero} provides a platform to test these myriad systems on the same task. Initial iterations of the challenge measured performance using the ABX discriminability criterion~\cite{schatz2013evaluating,schatz2014evaluating}. Since the 2019 challenge~\cite{dunbar2019zero}, there has been a shift to a synthesis-based evaluation scheme where the learned units are used to synthesize audio waveforms, and the quality of the synthesized waveforms - measured by human-evaluated Character Error Rate (CER) - is used as a proxy metric for the quality of the AUD. The rationale is that given a set of waveforms and their transcriptions, one should be able to train synthesis systems, and the more accurate the transcripts are, the less garbled the synthesis would be. The other relevant metric for us is the bitrate which measures the conciseness of the representation, i.e. if the AUD system incorrectly ascribes multiple units to the same phone, while the synthesis may not degrade much, the bitrate would increase.

\begin{figure*}[t]
    \begin{subfigure}[t]{\textwidth}
    \centering
    \includegraphics[scale=0.825]{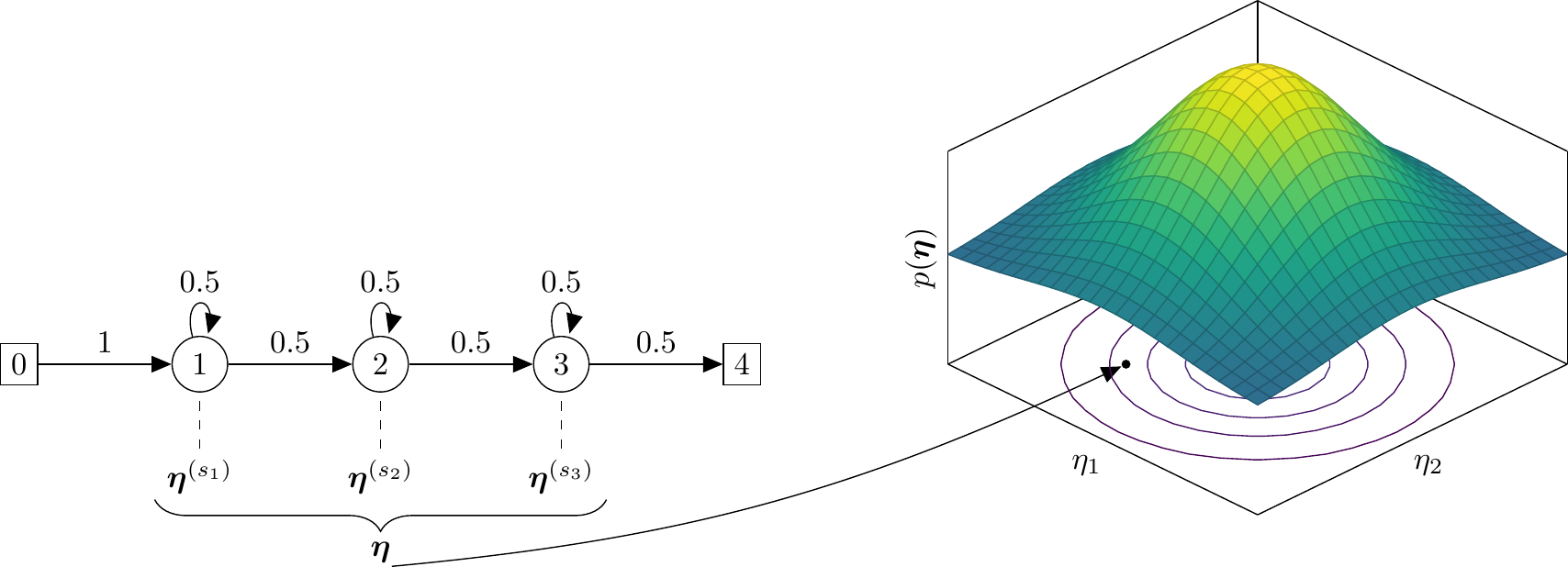}
    \caption{Model of an acoustic unit and its relation with the base measure $p(\Matrix{\eta})$ in the Dirichlet Process HMM AUD model \cite{lee2012nonparametric, ondel2016variational}.}
    \label{fig:hmm_unit}
\end{subfigure}

\begin{subfigure}[t]{\textwidth}
    \centering
    \includegraphics[scale=0.825]{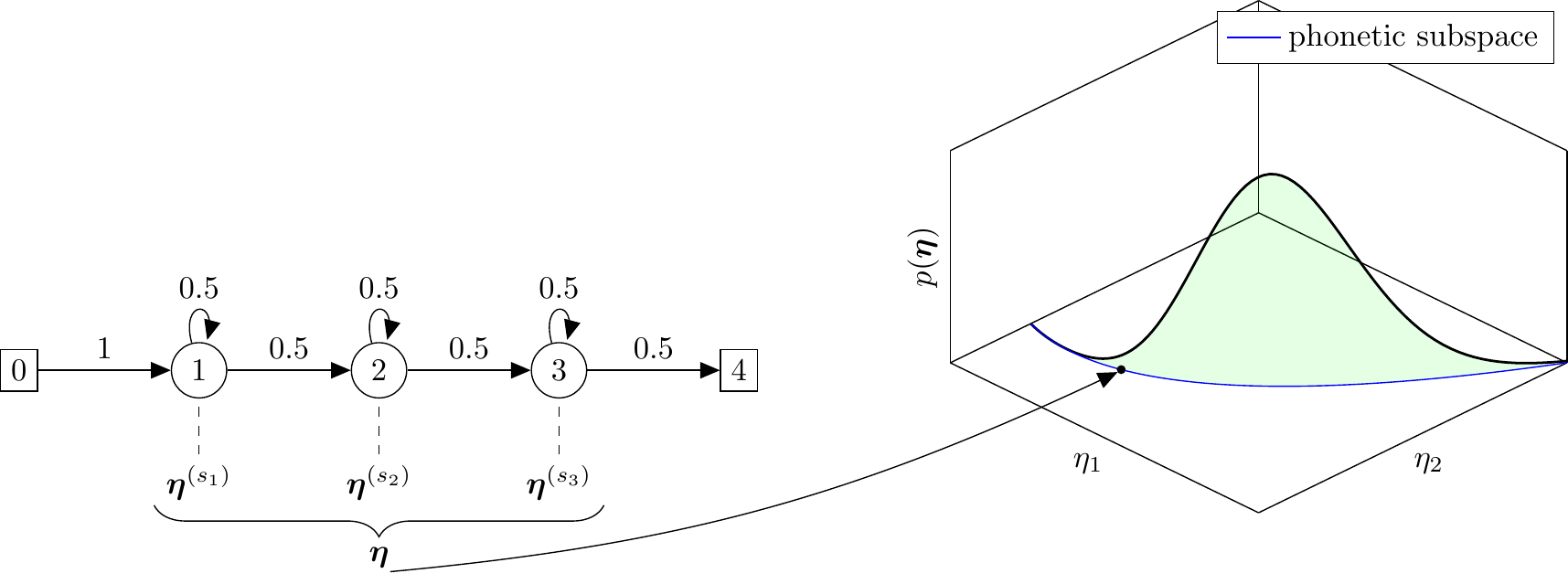}
    \caption{Model of an acoustic unit and its relation with the base measure $p(\Matrix{\eta})$ in the Dirichlet Process SHMM AUD model \cite{Ondel2019shmm}.}
    \label{fig:shmm_unit}
\end{subfigure}
\end{figure*}

\section{System}

\subsection{Bayesian Acoustic Unit Discovery}
The Bayesian approach to AUD involves maximizing the joint likelihood of the acoustic unit sequence $\Matrix{u}=(\Matrix{u}_1 \dots \Matrix{u}_N)$ and the set of \Lucas{acoustic unit} parameters $\Matrix{H} = \{ \Matrix{\eta}_1, \Matrix{\eta}_2, \dots \}$ conditioned on the observed data $\Matrix{X}=(\Matrix{x}_1, \dots \Matrix{x}_N)$:
\begin{align}
    p(\Matrix{u}, \Matrix{H} | \Matrix{X}) &\propto p(\Matrix{X} | \Matrix{u}, \Matrix{H}) p(\Matrix{u}, \Matrix{H})
    \label{eq:bayes},
\end{align}
where $\Matrix{\eta}_u$ is the vector of parameters \Lucas{of the probabilistic model} of acoustic unit $u$. \Lucas{In this work, we have used the non-parametric Bayesian AUD model proposed in~\cite{lee2012nonparametric, ondel2016variational} where} the likelihood of a speech segment given a specific unit $p(\Matrix{x}_{t_0}, \Matrix{x}_{t_1}, \dots, \Matrix{x}_{t_n}|u, \Matrix{\eta}_u)$ is modeled with a 3-state left-to-right HMM with parameters:
\begin{equation*}
    \Matrix{\eta}_u = \begin{pmatrix}
    \Matrix{\eta}_u^{1}\\
    \Matrix{\eta}_u^{2}\\
    \Matrix{\eta}_u^{3}
    \end{pmatrix},
\end{equation*}
and each state has emission probabilities modeled as a GMM with $K$ Gaussian \Lucas{distributions}:
\begin{equation*}
    \Matrix{\eta}_u^i = \begin{pmatrix} \Matrix{\mu}_u^{i,1} \\ \vdots \\ \Matrix{\mu}_u^{i,K} \\ \Vectorize(\Matrix{\Sigma}_u^{i,1}) \\ \vdots \\ \Vectorize(\Matrix{\Sigma}_u^{i,K}) \\ \pi_u^{i,1} \\ \vdots \\ \pi_u^{i,K} \end{pmatrix},
\end{equation*}
where $\Vectorize$ is the vectorize operator, $\Matrix{\pi}^i_u$, $\Matrix{\mu}^{i,j}_u$ and $\Matrix{\Sigma}^{i,j}_u$ are the mixing weights, the mean and covariance matrix of the $i$th HMM state and the $j$th Gaussian component of the acoustic unit $u$. Note that in this case, $\Matrix{\eta}_u$ is the supervector formed by concatenating all the elements of the means, covariance matrices and mixture weights for all three states of $u$ \Lucas{and can be interpreted as the embedding of an acoustic unit in a high-dimensional space.}

\Lucas{Following~\cite{lee2012nonparametric,ondel2016variational}}, the joint prior over units and parameters further factorizes into:
\begin{align}
    p(\Matrix{u}, \Matrix{H}) &= p(\Matrix{u}|\Matrix{H})\prod_{i=1}^{\infty}p(\Matrix{\eta}_i),
\end{align}
\Lucas{where} the conditional prior over units $p(\Matrix{u}|\Matrix{H})$ has a Dirichlet process prior with base measure $p(\Matrix{\eta})$.

The base measure defines \Lucas{\textit{a priori} which sound---represented as an HMM parameter vector $\Matrix{\eta}$---is likely to be selected as an acoustic units by the clustering process}. In \Lucas{prior works} \cite{lee2012nonparametric, ondel2016variational, ondel2018bayesian, ondel2017bayesian}, this base measure is \Lucas{set as a combination of Normal-Wishart and Dirichlet distributions}. \Lucas{This choice of base measure was mostly driven by mathematical convenience rather than an educated guess about which sound is a plausible acoustic unit.} Figure~\ref{fig:hmm_unit} depicts the result of this assumption in a 2D parameter space (note that we use 2D for visualization purposes; the actual parameter dimensionality is much higher). The probability is quite diffuse in the parameter space, which means that \textit{a priori}, the units can model any sounds (including non-human ones) and \Lucas{potentially irrelevant} sources of variability \Lucas{(speaker, channels, emotion, ...)}.



\subsection{Subspace HMM}
\label{sec:shmm}

\Lucas{The Subspace HMM (SHMM) introduced in \cite{Ondel2019shmm} improved over the previous non-parametric HMM based AUD model by defining a more refined base measure which narrows the search of acoustic units within a subspace of the total parameters space. In this framework, each acoustic unit embedding is constrained to live in a low-dimensional manifold:
\begin{equation}
    \Matrix{\eta}_u = f(\Matrix{W} \Matrix{h}_u + \Matrix{b}),
\end{equation}
where  $f$ is a differentiable function, $\Matrix{h}_u$ is a low-dimensional embedding of the acoustic unit $u$ and $\Matrix{W}$ and $\Matrix{b}$ are the parameters of the subspace. In this work, the columns of $\Matrix{W}$ are assumed to span the subspace containing the phonetic variability.
} As shown in Figure~\ref{fig:shmm_unit}, the SHMM modifies the base measure $p(\Matrix{\eta})$ so that \Lucas{all} the probability mass is \Lucas{contained} in a lower dimensional manifold of the total parameter space (the blue curve in the figure). \Lucas{Furthermore, the SHMM incorporates the following priors over the subspace parameters:} 
\begin{eqnarray}
    W_{r,c} &\sim& \mathcal{N}(0, 1) \\
    \mathbf{b} &\sim& \mathcal{N}(\mathbf{0}, \mathbf{I}) \\
    \mathbf{h}_u &\sim& \mathcal{N}(\mathbf{0}, \mathbf{I}),
\end{eqnarray}
\Lucas{where $W_{r,c}$ is element at the $r$th row and $c$th column of matrix $\Matrix{W}$. Finally, the function $f$ ensures that the embedding $\Matrix{\eta}$ lives in the HMM parameter space and it is defined as:}
\begin{eqnarray}
    \pi_u^{i,j} &=& \frac{\exp\{\Matrix{W}^i_{\pi} \Matrix{h}_u + \Matrix{b}^i_{\pi}\}_j }{1 + \sum_{k=1}^{K-1}\exp\{\Matrix{W}^i_{\pi} \Matrix{h}_u + \Matrix{b}^i_{\pi} \}_k} \\
    \Matrix{\mu}_u^{i, j} &=& \Matrix{W}^i_{\mu} \Matrix{h}_u + \Matrix{b}^i_{\mu}  \\
    \Matrix{\Sigma}_u^{i, j} &=& \Diag (\exp\{ \Matrix{W}^i_{\Sigma} \Matrix{h}_u + \Matrix{b}^i_{\Sigma}  \}),
\end{eqnarray}
\Lucas{where $\exp$ is the element-wise exponential function and $\exp \{...\}_j$ is the $j$th element of the resulting vector. The matrix $\Matrix{W}^i_{\pi}$ is the subset of rows of the matrix $\Matrix{W}$ which is assigned to the mixing weights $\Matrix{\pi}^i$ of the $i$th HMM state. The matrices $\Matrix{W}^{i,j}_{\mu}$ and $\Matrix{W}^{i,j}_{\Sigma}$ are similarly defined for the mean and covariance matrix of the $j$th Gaussian distribution of $i$th HMM state.}

\subsection{Model training}
The SHMM is trained by optimizing the variational lower bound on the posterior $p(\Matrix{u}, \Matrix{H}|\Matrix{X})$. The procedure can be split into two parts: (i) optimizing the \Lucas{variational posterior of the }subspace parameters $q(\Matrix{W}, \Matrix{b})$, and (ii) acoustic unit discovery, i.e. optimizing the \Lucas{variational posteriors} $q(\Matrix{h}_1), q(\Matrix{h}_2), \dots$. Although both steps can be combined for a target language, this defeats the motivation of the SHMM which is to use the subspace to constrain the embeddings.

In training the subspace, we need the knowledge of the phones beforehand in order to ensure that the subspace models phonetic variability, but this is the entire goal of acoustic unit discovery. This dilemma is resolved by noting that there is considerable phonetic overlap across many languages. Therefore, the target language phonetic subspace is approximated by a phonetic subspace learned from other \textit{source languages} with annotated data (along with their corresponding unit embeddings and parameters).  This can be seen as imposing an educated prior on the units of a language. The base measure $p(\Matrix{\eta})$ is now modified to be:
\begin{align}
    \Matrix{\eta} \sim p(\Matrix{\eta}|\mathcal{X}_p, \mathcal{U}_p),
\end{align}
where $\mathcal{X}_p$ and $\mathcal{U}_p$ are the utterances and phonetic transcriptions of the source languages.

For the acoustic unit discovery in the unlabeled target language, the same variational optimization is done, except now the variational posterior over the subspace parameters is \Lucas{held fixed} and only the variational posterior of the embeddings $q(\Matrix{h}_1)q(\Matrix{h}_2) \dots$ are learned. Another key difference is that since no annotations are available, the HMM inference graph is no longer the forced-alignment graph but a ``pseudo-phone" loop~\cite{ondel2016variational}.

The optimization in both cases involves a mix of the forward-backward algorithm~\cite{rabiner1989tutorial} for the HMM latent parameters and a gradient ascent with the reparametrization trick~\cite{kingma2013auto} for $q(\Matrix{W})q(\Matrix{b})q(\Matrix{h}_1)q(\Matrix{h}_2)\dots$. A more in-depth treatment of the procedure can be found in the original SHMM paper~\cite{Ondel2019shmm}. Finally, the code for training the SHMM can be found here: https://github.com/beer-asr/beer.

\subsection{Decoding units}
We obtain our representation by Viterbi decoding on the HMM to find the sequence of units $\Matrix{u}^*$ such that:
\begin{align}
    \Matrix{u}^* &= \argmax_{\Matrix{u}}q^*(\Matrix{u})\\
    q^*(\Matrix{u}) &= \argmax_{q(\Matrix{u})} \langle \ln \frac{ p(\Matrix{X} | \Matrix{u}, \Matrix{H})^{\alpha}p(\Matrix{u},\Matrix{H})}{q(\Matrix{u})q(\Matrix{H})} \rangle_{q(\Matrix{u})q(\Matrix{H})},
    \label{eq:bayes_decode}
\end{align}
where $\alpha$ is a weighting factor called the acoustic scale used in ASR systems to correct some of the modeling inaccuracies of the HMM. Unless otherwise specified, we set the acoustic scale to 1.

\section{Experiments}
\subsection{Data and features}

We tested our system on the two languages provided as part of the challenge: English and the surprise language, both of which were also used for the 2019 challenge. The surprise language data is a standard Indonesian corpus collected in~\cite{indonesian1,indonesian2}. Each of these corpora has three subsets:
\begin{enumerate}
    \item Unit training set: The set of data used for training the acoustic units. This corpus totals about 15 hours per language.
    \item Voice training set: Speech collection for each target speaker of the synthesis. For English, there are two target speakers with about 2 hours each. For the surprise language, there is one target speaker with about 90 minutes.
    \item Test set: The set of utterances (about 30 minutes per language) to be decoded for scoring. Some of these utterances are also used for syntheses.
\end{enumerate}

To train the approximate subspace, we use annotated data from the GlobalPhone corpus~\cite{schultz2002globalphone}: we use subsets from the German (2.7 hours), Spanish (4.7 hours), Polish (3.4 hours) and French (3.8 hours) sets of the corpus.

The input to the system are 13-dimensional MFCC features along with their first and second derivatives with per utterance mean normalization applied.

\subsection{Evaluation metrics}
In keeping with the goal of the challenge as `TTS-without-T', we evaluate our systems mainly on synthesis accuracy, specifically the character error rate (CER) of the synthesized waveforms as transcribed by human evaluators. This metric is of course tied with the bitrate of the representations, as a system with very high bitrate (say MFCC) should conceptually have very good synthesis and low character error rate. Moreover, our goal is to do phone discovery, having a low bit rate is quite essential since we know that the phone sets of each language are quite limited.

The challenge provides two additional subjective metrics. The first is the Mean Opinion Score (MOS) which evaluates the quality of the synthesis. The other metric is the similarity score which measures the voice similarity to the target speaker, and somewhat measures the disentanglement of speaker information from the learned representation. In addition to these metrics, the challenge also features the ABX error rate, which measures the likelihood that for a given triphone example, its representation as learned by the system is farther from another example of the same triphone than it is from an example of another triphone by some pre-specified distance metric.

\subsection{Results and discussions}
The baseline system of the challenge uses the Bayesian HMM from our previous work~\cite{ondel2016variational}. The topline is a basic HMM-GMM trained with supervision. We submitted two systems based on the SHMM described in section~\ref{sec:shmm} with different truncation parameters (for details about the truncation parameter, see~\cite{blei2006variational}). We use the Levenshtein distance to compute the ABX.

We train the SHMM with a $100$-dimensional subspace, $4$ Gaussians per state and a concentration parameter of 50. For synthesis, we use Merlin-based synthesizer~\cite{Wu+2016} used for both the baseline and topline.

Tables~\ref{tab:res_eng}~and~\ref{tab:res_sur} show the challenge leaderboard\footnote{https://zerospeech.com/2020/results.html} results on the English and surprise language test sets respectively. The baseline and topline results from 2019 are also included. In both cases, the same system outputs (hence the identical ABX and bitrates) are sent for another round  of human evaluations. This leads to some variation in the results, especially in the surprise language where the CER is raised by 5\% for both the baseline and topline in the 2020 evaluations compared to the 2019 evaluations.

\begin{table}[t]
\centering
\caption{English results. X75 and X100 refer to the subspace model truncated at 75 and 100 units respectively. B-2019 and B-2020 refer to the baseline results from 2019 and 2020 respectively. Similarly T-2019 and T-2020 refer to the topline results from 2019 and 2020 respectively.}
\label{tab:res_eng}
\begin{tabular}{lccccc}
\hline
         & \multicolumn{1}{l}{MOS} & \multicolumn{1}{l}{CER} & \multicolumn{1}{l}{Similarity} & \multicolumn{1}{l}{ABX} & \multicolumn{1}{l}{Bitrate} \\ \hline
B-2019 & \textbf{2.5}                     & \textbf{0.75}                     & 2.97                            & \textbf{35.63}                    & 71.98                        \\
B-2020 & 2.14                     & 0.77                     & 2.98                            & \textbf{35.63}                    & 71.98                        \\

X75      & 2.12                     & 0.76                     & 2.91                            & 36.83                    & \textbf{58.98}                        \\
X100     & 1.99                     & \textbf{0.75}                     & \textbf{3.18}                            & 36.63                    & 64.16                        \\ \hline
T-2019 & 2.77                     & 0.44                     & 2.99                            & 29.85                    & 37.73                        \\
T-2020 & 2.52                     & 0.43                     & 3.10                             & 29.85                    & 37.73                        \\ \hline
\end{tabular}
\end{table}

\begin{table}[t]
\centering
\caption{Surprise language results}
\label{tab:res_sur}
\begin{tabular}{lccccc}
\hline
         & \multicolumn{1}{l}{MOS} & \multicolumn{1}{l}{CER} & \multicolumn{1}{l}{Similarity} & \multicolumn{1}{l}{ABX} & \multicolumn{1}{l}{Bitrate} \\ \hline
B-2019 & 2.07                     & \textbf{0.62}                     & 3.41                            & \textbf{27.46}                    & 74.55                        \\
B-2020 & 2.23                     & 0.67                     & 3.26                            & \textbf{27.46}                    & 74.55                        \\
X75      & \textbf{2.49}                     & \textbf{0.62}                     & \textbf{3.43}                            & 33.66                    & \textbf{65.76}                        \\
X100     & 2.37                     & \textbf{0.62}                     & 3.14                            & 33.99                    & 72.41                        \\ \hline
T-2019 & 3.92                     & 0.28                     & 3.95                            & 16.09                    & 35.2                         \\
T-2020 & 3.49                     & 0.33                     & 3.77                            & 16.09                    & 35.2                         \\ \hline

\end{tabular}
\end{table}

We submitted the SHMM system with truncation parameters of 75 and 100 units. While using fewer units leads to a lower bitrate, the character error rate stays the same.

Compared to the baseline, we improve on the surprise language baseline CER by 5\%. However, this improved result is identical to the baseline CER from the 2019 evaluation so it is likely within the error margin of the human evaluators. In any case, we are able to at least match the baseline CER with considerably lower bitrate: up to 13 bits-per-second (bps) improvement on English and 9 bps on the surprise language. We also found that decreasing the concentration parameter of the Dirichlet process (from 50 to 1) reduced the bitrate by a further 5 bps while maintaining the same ABX on the English test set, but since this was done post-eval, we have no human evaluation scores for the resulting synthesis.

While we do not focus on tuning our system for the ABX, in our preliminary experiments, we found that the metric favors our denser representations. In particular, using the posteriors obtained from the forward-backward algorithm as our embeddings improves the ABX (computed with the DTW+KL-divergence as opposed to the Levenshtein distance used for the transcripts). Furthermore, by reducing the acoustic scale in~\eqref{eq:bayes_decode}, we can make the posteriors fuzzier and  so decrease the ABX at the cost of higher bitrate. We do not make use of these in our final submission since our goal is the discovery of phones which are discrete.
\begin{figure}[t]
    \centering
    \resizebox{0.47\textwidth}{!}{%
        \begin{tikzpicture}
	\begin{axis}[
	    legend style={at={(0.7,0.5)},anchor=west,draw=none},
		xlabel=Acoustic scale ($\alpha$),
		ylabel=ABX]
    \addplot[dashed,color=blue,mark=*] coordinates {
        (0.2,19.759)
        (0.5,23.038)
        (1.0,27.215)
        (1.5,29.396)
        (2.0,30.537)
        (2.5,31.544)
        (5.0,32.47)
        (9.0,33.165)};
        \label{plot_ac_abx}
        \addlegendimage{/pgfplots/refstyle=plot_ac_abx}\addlegendentry{ABX}
	\end{axis}
	\begin{axis}[
	    legend style={at={(0.7,0.4)},anchor=west,draw=none},
    	ylabel=Bitrate (bps),
    	axis y line*=right,
    	axis x line=none]
	\addplot[color=red,mark=x] coordinates{
        (0.2, 1281)
        (0.5, 1203)
        (1.0, 1085)
        (1.5, 994)
        (2.0, 922)
        (2.5, 869)
        (5.0, 726)
        (9.0, 640)
        };
        \label{plot_ac_bitrate}
        \addlegendimage{/pgfplots/refstyle=plot_ac_bitrate}\addlegendentry{Bitrate}
	\end{axis}

        \end{tikzpicture}
    }
    \label{fig:abx}
\end{figure}

\section{Conclusions}
In this paper, we have described our submission to the Zerospeech 2020 challenge where we have used the non-parametric Bayesian SHMM model for acoustic unit discovery. This model differs from the original HMM based system by having a constrained base measure---a distribution defining \textit{a~priori} which sound is a potential acoustic unit candidate---forcing the discovered acoustic unit embeddings to live in a low dimensional manifold previously estimated on annotated data from other languages. Compared to the baseline, we are able to get similar character error rates with much lower bitrates. In addition, we observed that, by taking the posterior over the acoustic units rather than the Viterbi path, one can achieve a very low ABX score at the expense of a much higher bitrate.



\bibliographystyle{IEEEtran}

\bibliography{mybib}


\end{document}